\documentclass[a4paper,11pt]{article}
\usepackage{jinstpub} 

\title{\boldmath Calibration of an Irradiated Prototype for the EIC Zero-Degree Calorimeter }

\author[a]{Weibin Zhang,}
\author[a]{Xilin Liang,}
\author[a]{Sean Preins,}
\author[a]{Miguel Arratia}
\affiliation[a]{Department of Physics and Astronomy, University of California, Riverside, CA 92521, USA}

\emailAdd{miguel.arratia@ucr.edu}

\abstract{We study the response of a prototype Zero-Degree Calorimeter (ZDC) detector to irradiation equivalent to 10$^{11}$ 1-MeV $n_{\text{eq}}/\text{cm}^{2}$, which matches the expected exposure after one year of operation at full nominal luminosity at the future Electron-Ion Collider (EIC). The prototype, which consists of 563 channels and represents about 10 percent of the final ZDC design in terms of both channel count and detector volume, was irradiated at the NASA Space Radiation Laboratory (NSRL) at Brookhaven National Laboratory (BNL) with proton beams. We demonstrate that, despite significant radiation damage to the SiPMs and non-uniform degradation across the detector volume, the detector can be successfully calibrated on a channel-by-channel basis using cosmic-ray data. The damage profile, similar to what is expected in the experiment, varies by an order of magnitude or more across the detector. Even for the most heavily damaged channels, the signal-to-noise ratio for a MIP signal remains above 5. This study provides a realistic test of the system's performance under irradiation. It complements previous SiPM-specific irradiation studies and will inform the future operation of the ZDC and other detectors that use SiPM-on-tile technology.}


\arxivnumber{2512.20852} 

\begin{document}
\maketitle
\flushbottom

\section{Introduction}
\label{sec:introduction}

The future Electron–Ion Collider (EIC) \cite{Accardi2016, ABDULKHALEK2022122447} to be constructed at Brookhaven National Laboratory (BNL) will be the first collider capable of delivering high-luminosity collisions of polarized electrons with polarized protons and light ions, as well as unpolarized heavy ions. The EIC will enable a broad program to study the spin and spatial structure of the nucleon and the emergent properties of nuclear matter in terms of quarks and gluons. To fully realize this program, the ePIC detector \cite{ePIC} has been designed as a comprehensive multi-purpose apparatus with specialized subsystems in the forward and backward regions.

Silicon photomultiplier (SiPM)-on-tile technology \cite{White_2023,refId0} has been widely adopted across ePIC due to its compactness, scalability, and excellent photon detection efficiency. It is employed in both central and forward calorimetry, where the expected annual radiation levels vary by orders of magnitude, from about $10^{9}$ to $10^{12}$ 1-MeV $n_{\text{eq}}/\text{cm}^{2}$ depending on detector position \cite{ABDULKHALEK2022122447}. Among these systems, the Zero-Degree Calorimeter (ZDC) \cite{Milton_2025} is particularly exposed, as it detects neutral particles emitted at very small angles with respect to the beamline. The ZDC provides essential capabilities for tagging spectator neutrons in electron–ion collisions and identifying diffractive processes, and thus must maintain stable operation under sustained high radiation doses.

This work continues a sequence of SiPM-on-tile calorimeter developments carried out by our group. The first-generation device, consisting of 10 layers and 40 channels, demonstrated the feasibility of the concept in a controlled beam test at Jefferson Lab \cite{instruments7040043}. A second-generation instrument, with 20 layers and 192 channels, was deployed in the STAR experiment at RHIC, where it provided the first operational validation of the technology under real collision and radiation conditions \cite{Zhang_2025}. The present third-generation prototype extends the program with 23 layers and 563 channels, representing approximately 10\% of the final ZDC design. The prototype was exposed to controlled irradiation at the NASA Space Radiation Laboratory (NSRL) at BNL, reaching a fluence of $10^{11}$ 1-MeV $n_{\text{eq}}/\text{cm}^{2}$, equivalent to one year of EIC operation at nominal luminosity. This provided a realistic, system-level irradiation test of the detector.

This paper studies pedestal and minimum ionizing particle (MIP) calibrations as diagnostics of SiPM degradation and detector stability under irradiation. Pedestal shifts and broadenings directly reflect radiation-induced dark current and noise evolution, while the MIP response provides a robust anchor for gain calibration and energy reconstruction. Together, these observables provide a comprehensive picture of SiPM performance as a function of radiation dose. By studying their evolution before and after irradiation, we demonstrate that despite substantial and non-uniform SiPM damage, channel-by-channel calibration with cosmic-ray data remains feasible and ensures adequate signal-to-noise performance. These results provide an essential benchmark for the long-term operation of the ePIC ZDC and offer broader insight for other ePIC subsystems based on similar technology.

This paper is organized as follows. Section~\ref{sec:prototype} describes the design and construction of the prototype and the irradiation setup. Section~\ref{sec:analysis} outlines the calibration methodology. Section~\ref{sec:results} presents the results of pedestal and MIP studies before and after irradiation. Section~\ref{sec:conclusion} discusses the implications of these findings for ZDC operation at the EIC, and provides a summary and outlook.
\section{Prototype}
\label{sec:prototype}

\begin{figure}
    \centering
    \includegraphics[width=0.58\linewidth]{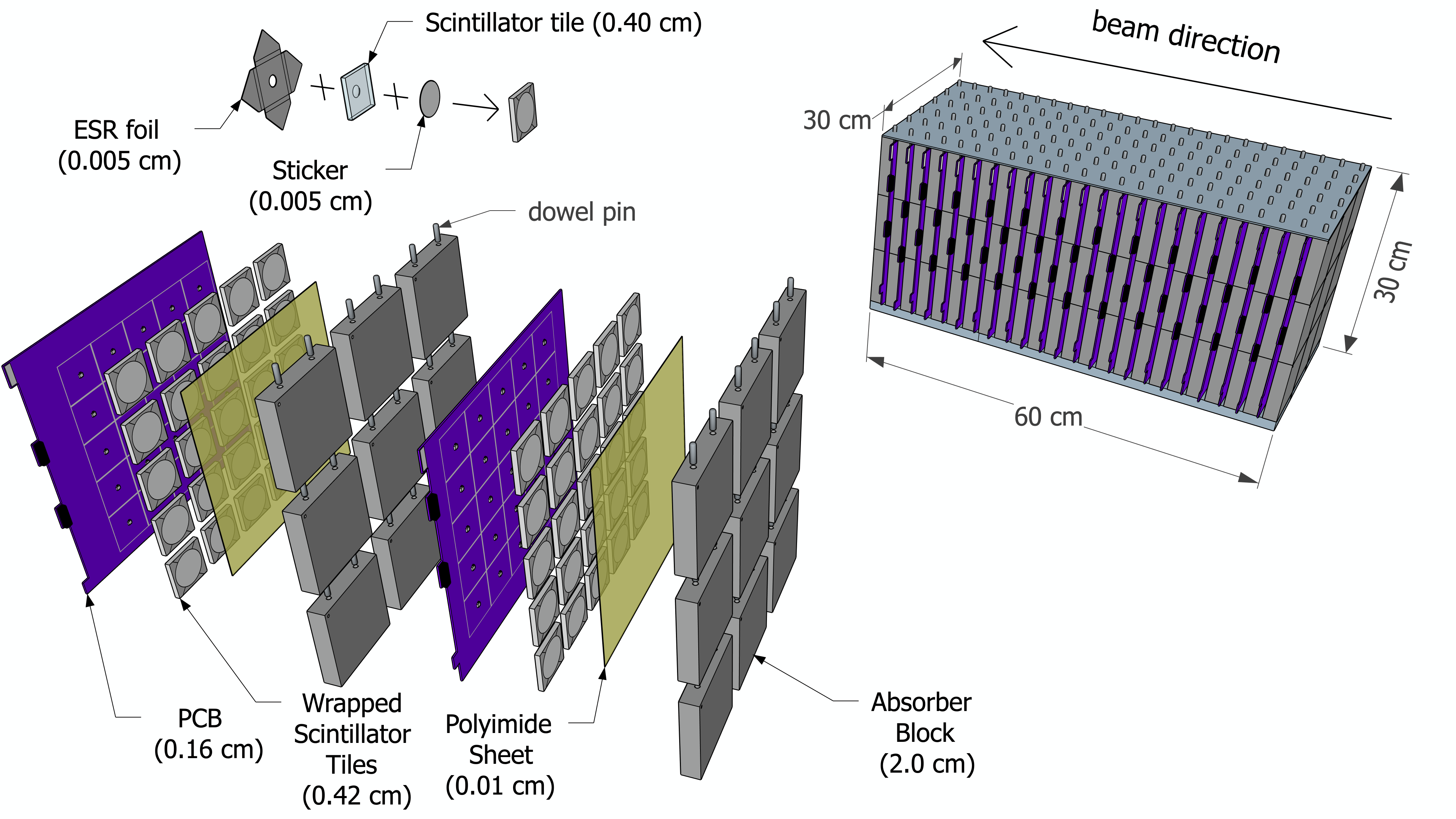}
    \includegraphics[width=0.41\linewidth]{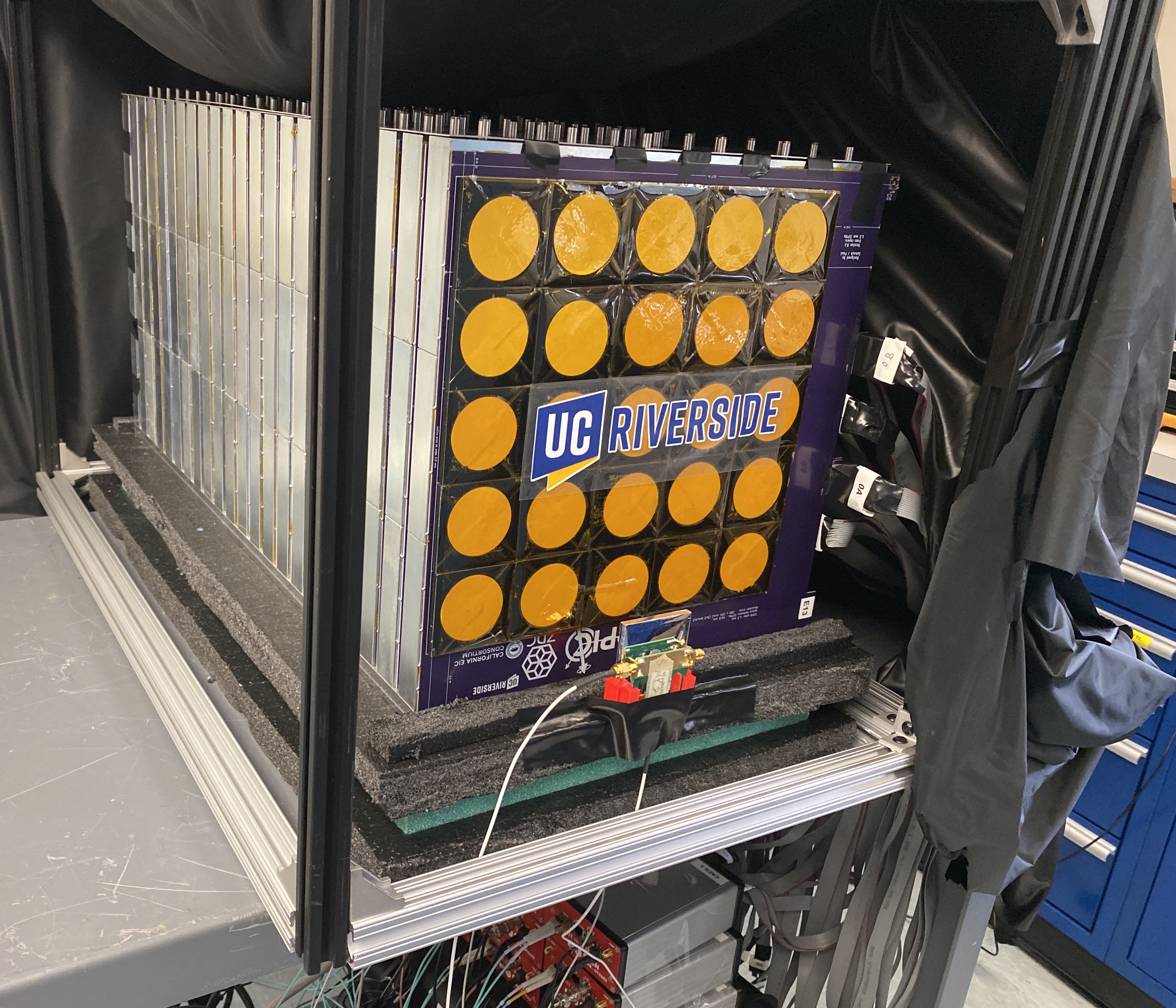}
    \caption{Schematic plot (left) and picture (right) of the ZDC prototype.}
    \label{fig:zdc}
\end{figure}

The prototype is a sampling calorimeter with 23 active layers interleaved with 22 absorber layers, as shown in figure~\ref{fig:zdc}. Each layer consists of a 2~cm-thick steel absorber, a scintillator-tile layer, and a custom PCB with SiPMs mounted directly on it. The absorber layers are built from a $3\times3$ array of steel blocks, while each active layer contains 25 square scintillator tiles arranged in a $5\times5$ grid. Odd and even layers are diagonally shifted relative to each other to create smaller effective cells and improve spatial resolution, using the HEXPLIT reconstruction algorithm~\cite{Paul_2024}.

The steel blocks were reused from the STAR FCS upgrade \cite{Wang_2024}. Each block is a square prism measuring $9.5\times9.5\times2~\mathrm{cm^3}$, with two pairs of blind holes on opposite faces (two per face) to accommodate dowel pins. The 2~cm thickness corresponds to $1.14$ radiation lengths ($X_0$) and approximately $0.5$ hadronic interaction lengths ($\lambda_I$).

Each active layer consists of 25 EJ-212 scintillating tiles \cite{EJ212} wrapped in reflective ESR foil. A central dimple on the back of each tile houses a Hamamatsu S14160-1315PS 1.3-mm SiPM, air-coupled to the tile. The tiles are adhered onto a custom PCB that routes signals to adapter connectors for readout.

Assembly began with three bottom steel plates with pre-drilled holes. The absorber blocks were then stacked layer by layer on top of the plates, up to three blocks high, and connected using dowel pins. An aluminum top plate, machined with the same hole pattern as the combined bottom plates, was placed on top of the absorber stack and secured with pins. The stacked absorber blocks created slots into which the PCB–scintillator assemblies were inserted from the side, with careful longitudinal alignment to maintain co-linearity with the absorber grid. The completed prototype measured approximately $30\times30\times60~\mathrm{cm^3}$. A frame made of 80-20 bars, wrapped with blackout fabric, covered the prototype to make it light-tight.

Each PCB is equipped with two adapter connectors, each supporting up to 13 channels. Custom 1~m ribbon cables link the adapters to CAEN FERS-5200 units (model DT5202) \cite{DT5202}, with compatible ports. Each CAEN unit provides five adapter ports; nine units are used to cover 22.5 layers, with the final layer reading out only half of its channels. In total, the system reads out 563 channels. 

The CAEN FERS system, based on the CITIROC-1A ASIC \cite{citiroc}, is adopted for prototyping and beam-test operation, while the final readout electronics for the EIC calorimeters, based on the EIC-custom CALOROC ASIC \cite{Dumas:2025dhq}, are still under development and not yet available. While this setup does not constitute a complete full-chain test, given the use of CITIROC-1A instead of CALOROC, it provides a first step toward validating the readout design, as similar performance is expected from the two ASICs and the signal-to-noise ratio is anticipated to be dominated by the SiPM dark current.

All CAEN units were connected, either in series or parallel, to a CAEN A3261 concentrator \cite{DT5215} via TDLink. The concentrator controls and synchronizes all CAEN units. The concentrator itself was connected to a laptop via USB and operated with the JANUS software \cite{Janus}.

\begin{figure}[h!]
    \centering
    \includegraphics[width=0.5\linewidth]{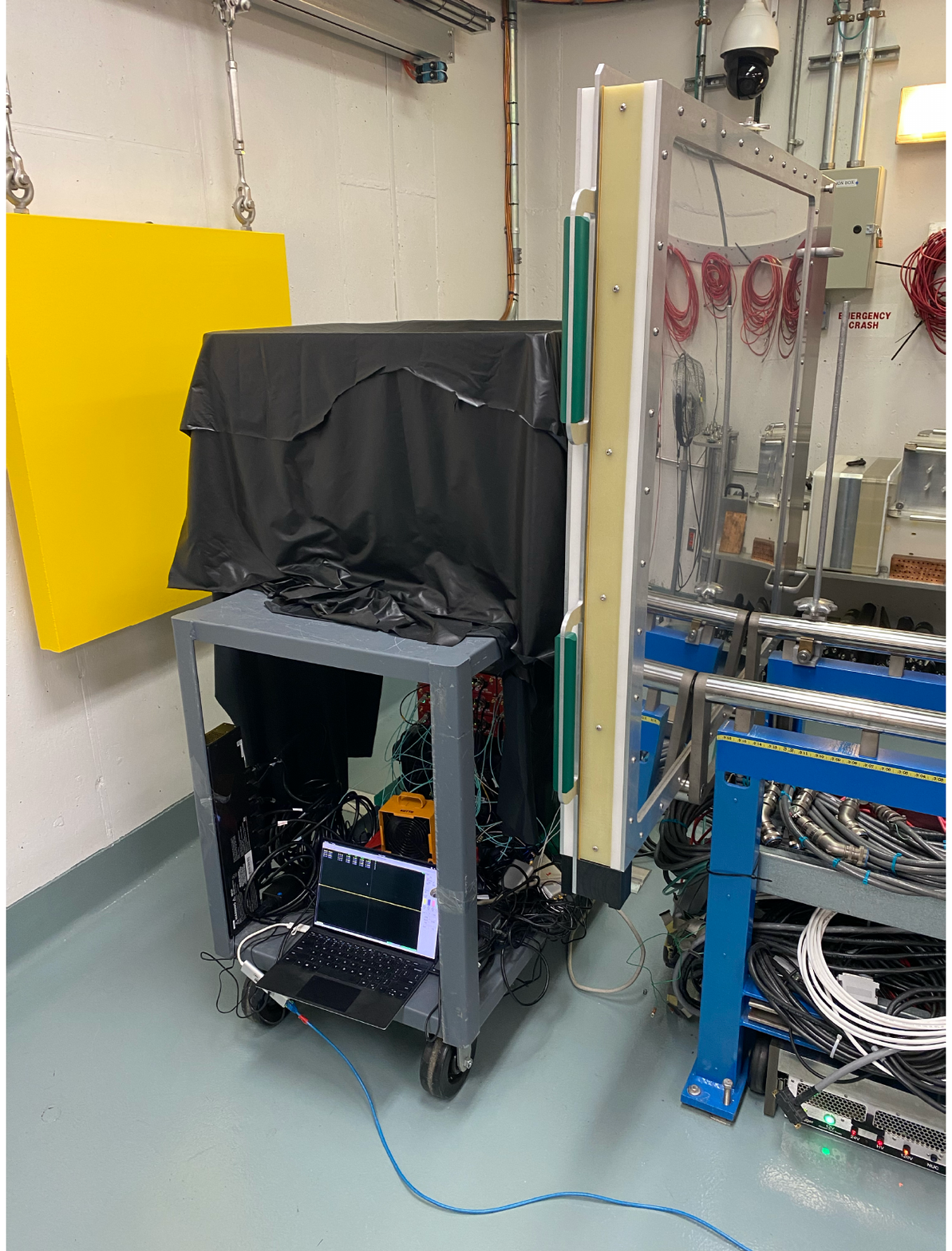}
    \caption{ZDC installed at the beam dump in the NSRL target room.}
    \label{fig:zdc_at_NSRL}
\end{figure}

The prototype was positioned at the beam dump in the NSRL target room, approximately aligned such that its geometric center was close to the nominal beam axis and its front face was oriented perpendicular to the beam trajectory, as shown in figure~\ref{fig:zdc_at_NSRL}. It was subsequently exposed to various high-intensity ion beams. 
Owing to the sampling calorimeter’s multi-layer geometry and the longitudinal attenuation of the beam, different layers and their corresponding SiPMs experienced varying radiation levels, enabling a systematic study of the SiPM response as a function of accumulated dose.

\section{Analysis}
\label{sec:analysis}

Data acquisition was performed using the Janus software package to configure and control the CAEN FERS-5200 readout units. The CAEN system digitizes each input signal through two parallel amplification paths, a high-gain and a low-gain channel, enabling an extended dynamic range and allowing simultaneous sensitivity to small signals and large energy deposits. Two trigger modes were employed for this study: PTRG (periodic trigger) mode and TLOGIC (trigger logic) mode. In PTRG mode, the system generated periodic triggers independent of detector activity, 
enabling unbiased sampling of the electronic response\footnote{The peak sensing feature of CITIROC-1A ASIC causes the measured pedestal to be slightly elevated and broadened relative to the true electronics baseline}. This mode was used for pedestal characterization. In TLOGIC mode, each CAEN unit was triggered independently based on a multiplicity logic of two or more coincident hits. This configuration was chosen to preferentially select cosmic-ray events traversing multiple tiles while suppressing random noise triggers.

Pedestal data were collected in PTRG mode with the beam off. For each channel, raw ADC distributions were accumulated over many triggers. These distributions were fitted with Gaussian functions, from which the mean and standard deviation were extracted and interpreted as the pedestal level and its width (electronic noise), respectively. Channels with unstable pedestals or anomalously high noise were excluded from further analysis. The pedestal values determined in this way were used as reference baselines for calibration.

MIP data were obtained with cosmic rays using the in situ prototype, positioned as during the beam test. Data were collected in TLOGIC mode, with each CAEN unit operating its own local coincidence trigger, and the SiPM bias voltage was set to 43~V both before and after irradiation. A unit would issue a trigger when two or more of its connected channels recorded signals above threshold within a predefined coincidence window of 100~ns, with the signal shaping time set to 25~ns. This two-hit coincidence reflected both the geometry of the instrumented region (each unit servicing at most two and a half PCBs) and the expectation that a single cosmic-ray muon would produce one hit per layer along its path.

For each channel, the raw ADC spectra exhibited two distinct peaks: a pedestal peak and a MIP peak. The high-gain (HG) setting was chosen sufficiently high to ensure these peaks were well separated. The MIP peak position was identified by inspection as the channel’s most probable value (MPV) for a minimum-ionizing particle. Subtracting the corresponding pedestal value from this MPV yielded the MIP calibration constant for that channel. This procedure provided a consistent response scale in MIP-equivalent units across all layers and channels.
\begin{figure}
    \centering
    \includegraphics[width=0.49\linewidth]{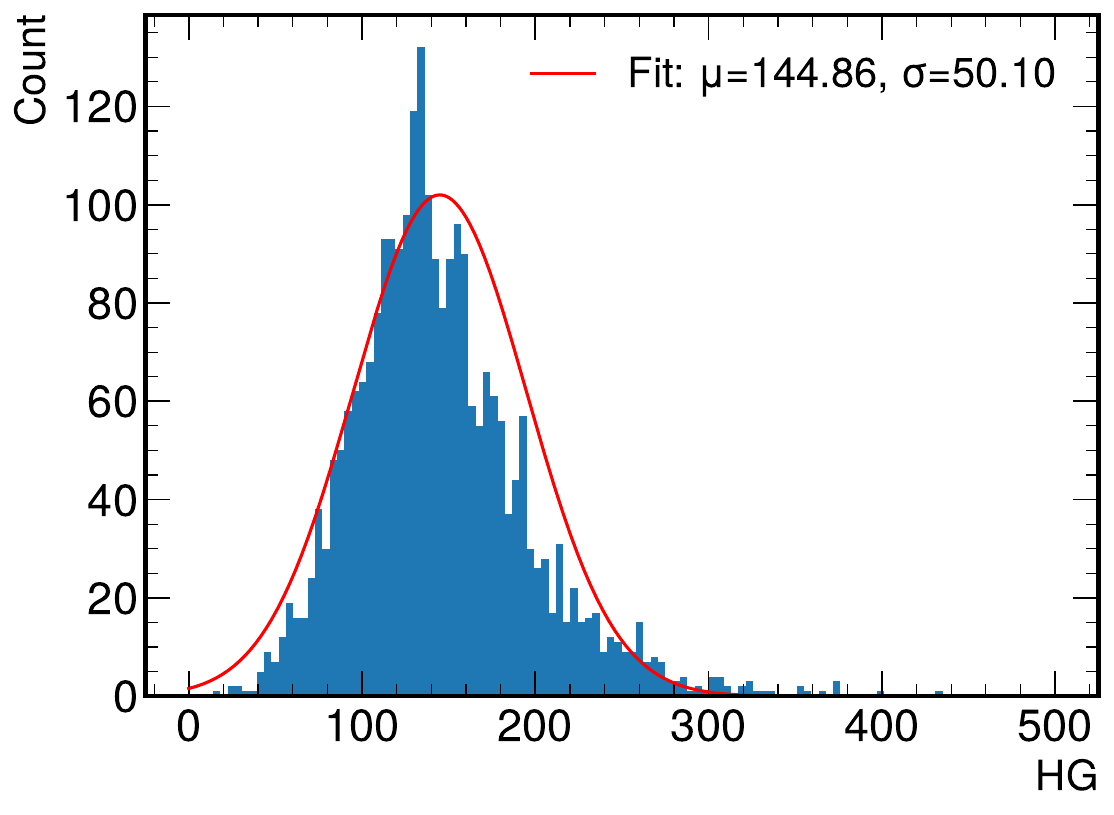}
    \includegraphics[width=0.49\linewidth]{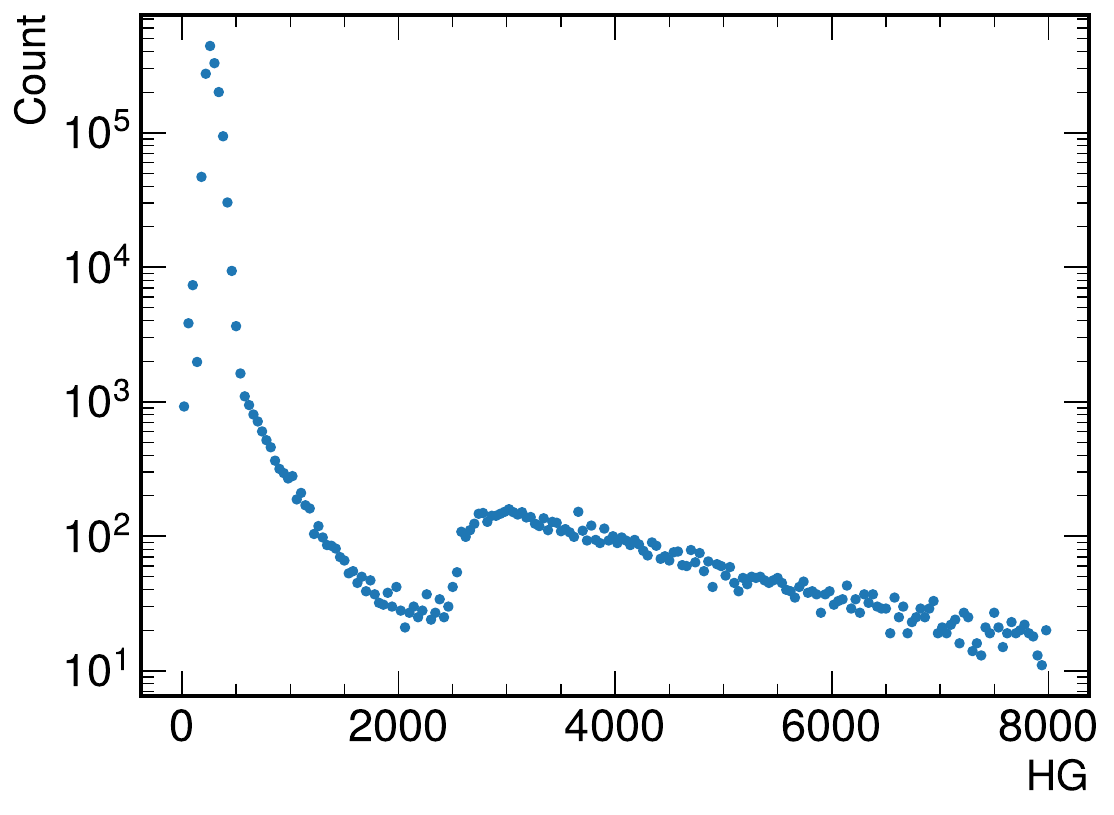}
    \caption{Left: a typical pedestal distribution fitted with a Gaussian function before irradiation; right:  a typical MIP ADC spectrum before irradiation.}
    \label{fig:example_spectra_before_irradiation}
\end{figure}

\section{Results}
\label{sec:results}

\subsection{Irradiation Profile of the Prototype}
\label{subsec:results_radiation}

\begin{figure}[h!]
    \centering
    \includegraphics[width=0.40\linewidth]{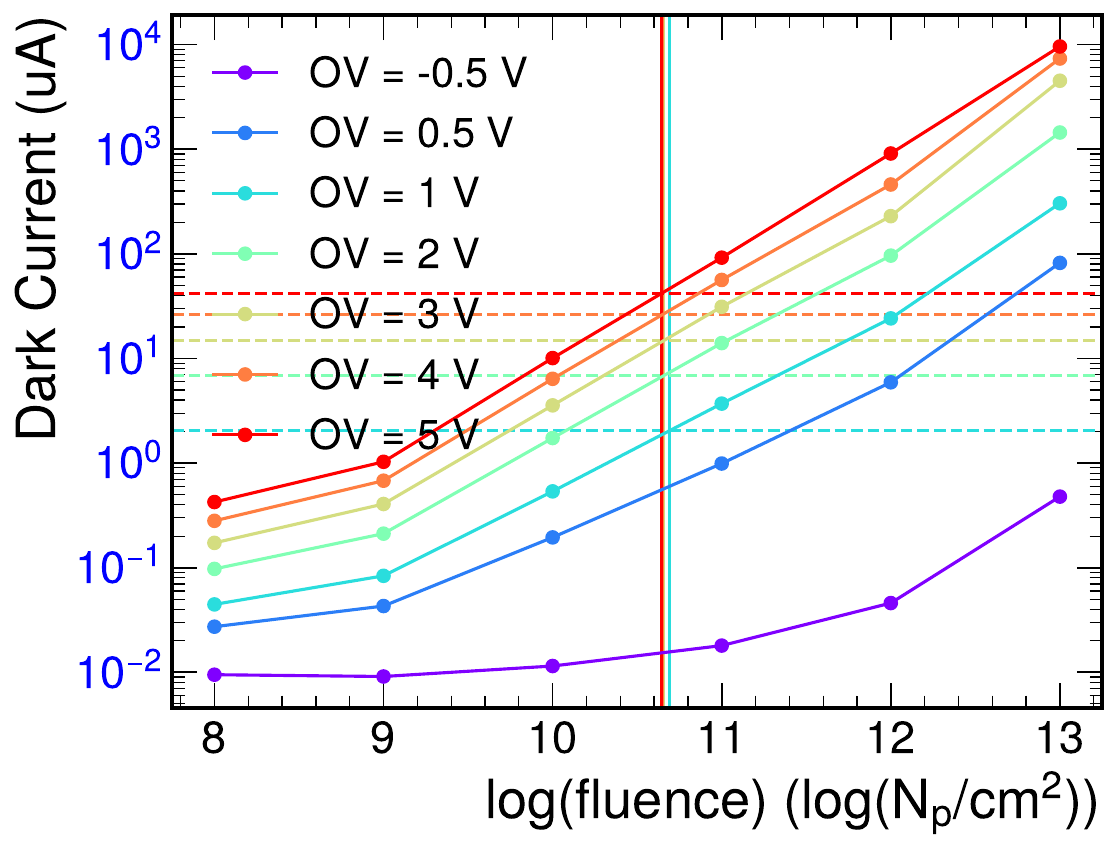}
    \includegraphics[width=0.59\linewidth]{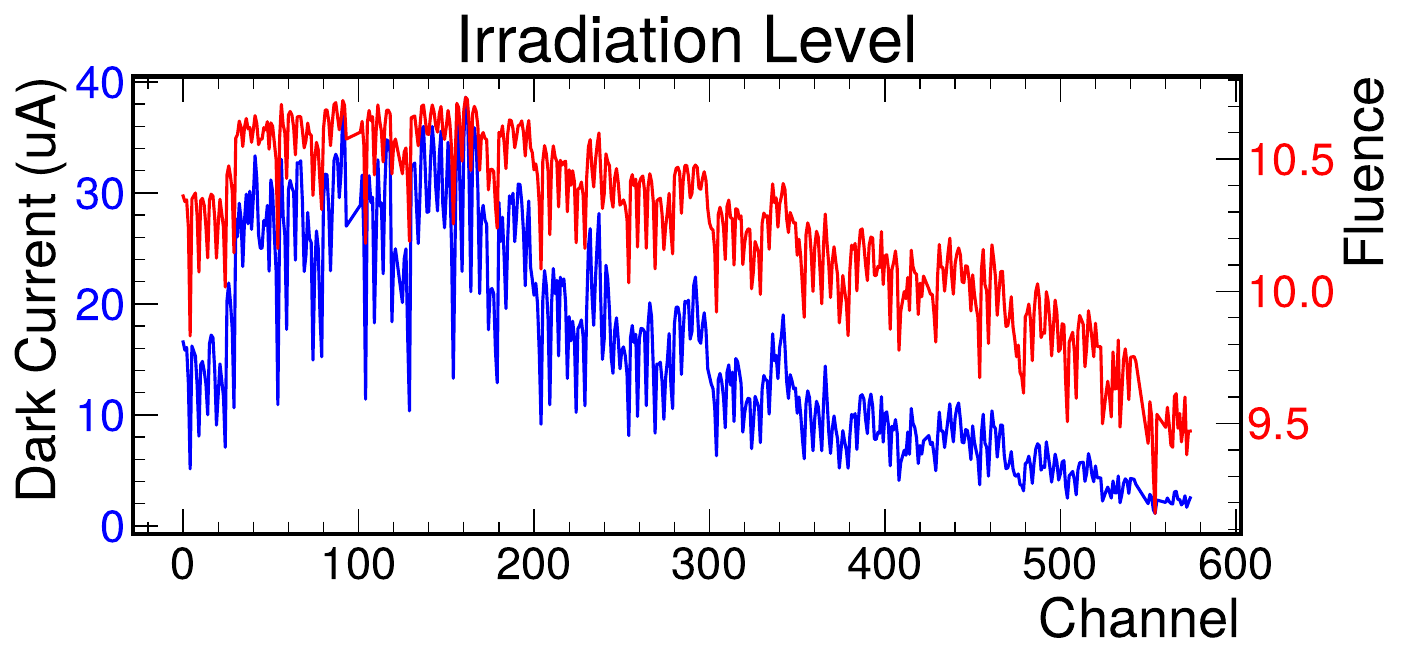}
    \caption{Left: Example dark current measurement for a single SiPM (Hamamatsu S14160 1315PS) channel at multiple bias voltages, mapped onto benchmark curves to infer the radiation fluence. Right: Inferred radiation fluence for all SiPM channels in the prototype, showing the variation across layers. The right axis utilizes a log scale.}
    \label{fig:irradiation}
\end{figure}

The radiation dose received by each SiPM in the prototype was inferred from post-irradiation dark current measurements.\footnote{The room temperature annealing effect is negligible based on our previous study \cite{huang2025}.} An example for a single channel, measured across several bias voltages, is shown in the left panel of figure~\ref{fig:irradiation}. Using benchmark curves obtained from previous irradiation studies \cite{huang2025}, which map the relationship between SiPM dark current and accumulated proton fluence, the corresponding radiation level for this channel was determined by intersecting the measured dark current on the benchmark curve. The inferred fluence is consistent across different bias voltages, demonstrating the robustness of the method.

Applying this procedure to all channels, the right panel of figure~\ref{fig:irradiation} shows the inferred radiation level for the entire prototype. The multi-layer geometry of the sampling calorimeter leads to a strong variation in radiation exposure, with front layers receiving the highest fluence, up to $10^{10.8}$ 64-MeV proton/cm$^2$ ($10^{10.98}$ 1-MeV $n_\text{eq}/\text{cm}^{2}$), and rear layers the lowest, about $10^{9.4}$ 64-MeV proton/cm$^2$ ($10^{9.58}$ 1-MeV $n_\text{eq}/\text{cm}^{2}$). This nearly order-of-magnitude variation across the detector provides a realistic test environment for studying SiPM performance under differential radiation levels.

For context, simulations indicate that under nominal conditions, an integrated luminosity of 1~fb$^{-1}$ yields a typical fluence of approximately $10^{9}$ 1-MeV $n_\text{eq}/\text{cm}^2$ for the ZDC \cite{ePICFluence}. Scaling linearly to 100~fb$^{-1}$, corresponding to one year of running at full nominal luminosity, the annual fluence reaches $10^{11}$~1-MeV $n_\text{eq}/\text{cm}^2$, matching the radiation dose applied to the prototype in the NSRL test. This demonstrates that the irradiation study realistically reproduces the operational radiation environment expected for the final detector, providing a solid basis for evaluating SiPM response and calibration stability.


\subsection{Pedestal Calibration}
\label{subsec:results_pedestal}

\begin{figure}[h!]
    \centering
    \includegraphics[width=0.49\linewidth]{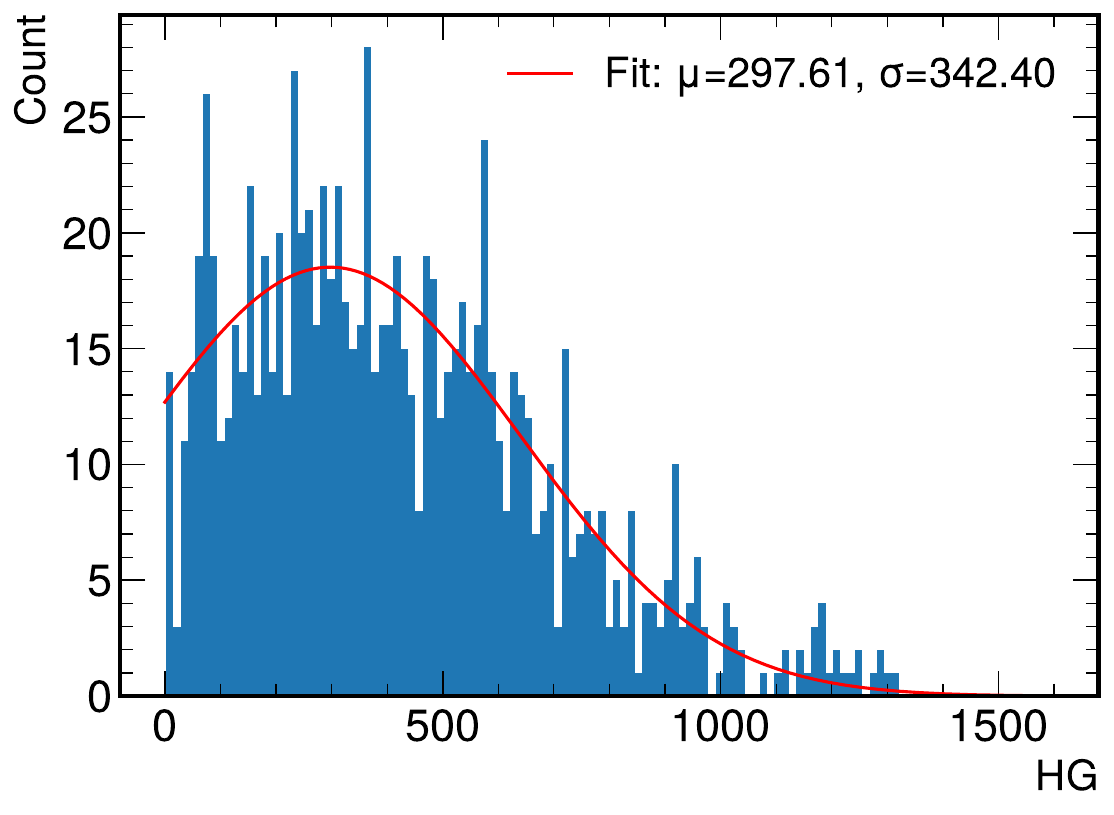}
    \includegraphics[width=0.49\linewidth]{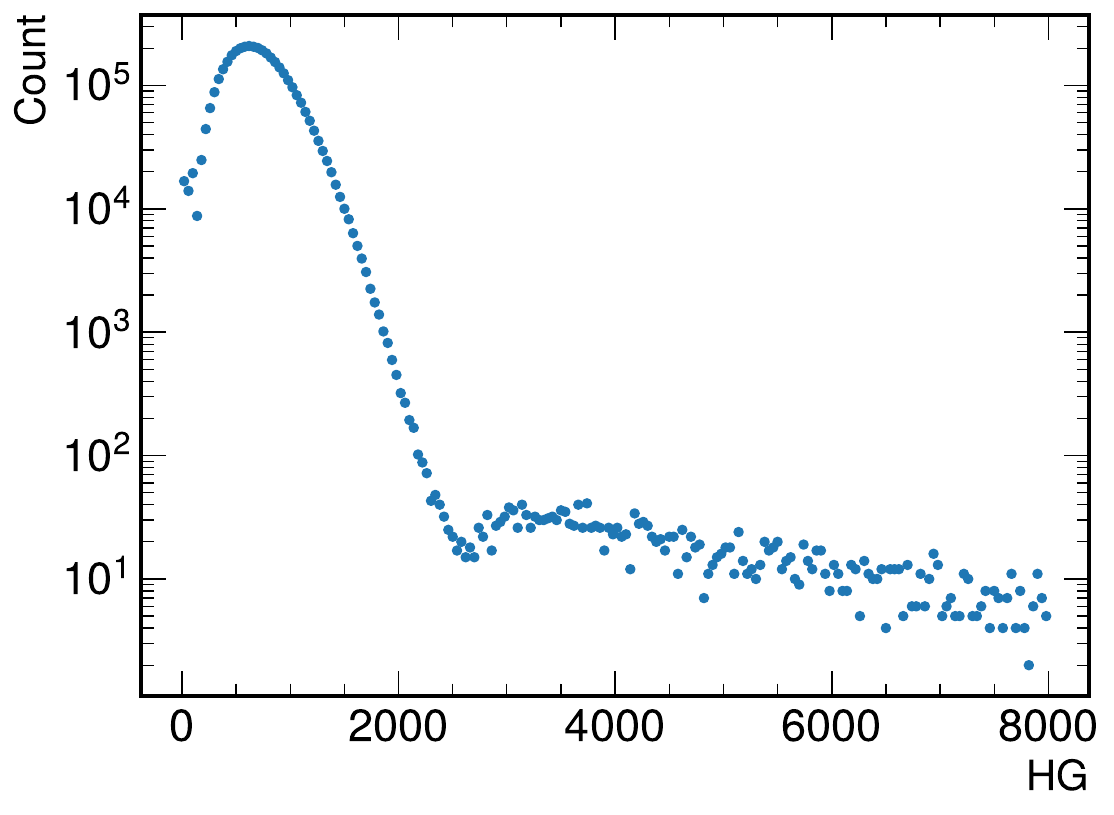}
    \caption{Representative pedestal (left) and MIP (right) spectra after irradiation. The corresponding pre-irradiation spectrum is shown in figure~\ref{fig:example_spectra_before_irradiation}.}
    \label{fig:example_spectra_after_irradiation}
\end{figure}

\begin{figure}[h!]
    \centering
    \includegraphics[width=0.45\linewidth]{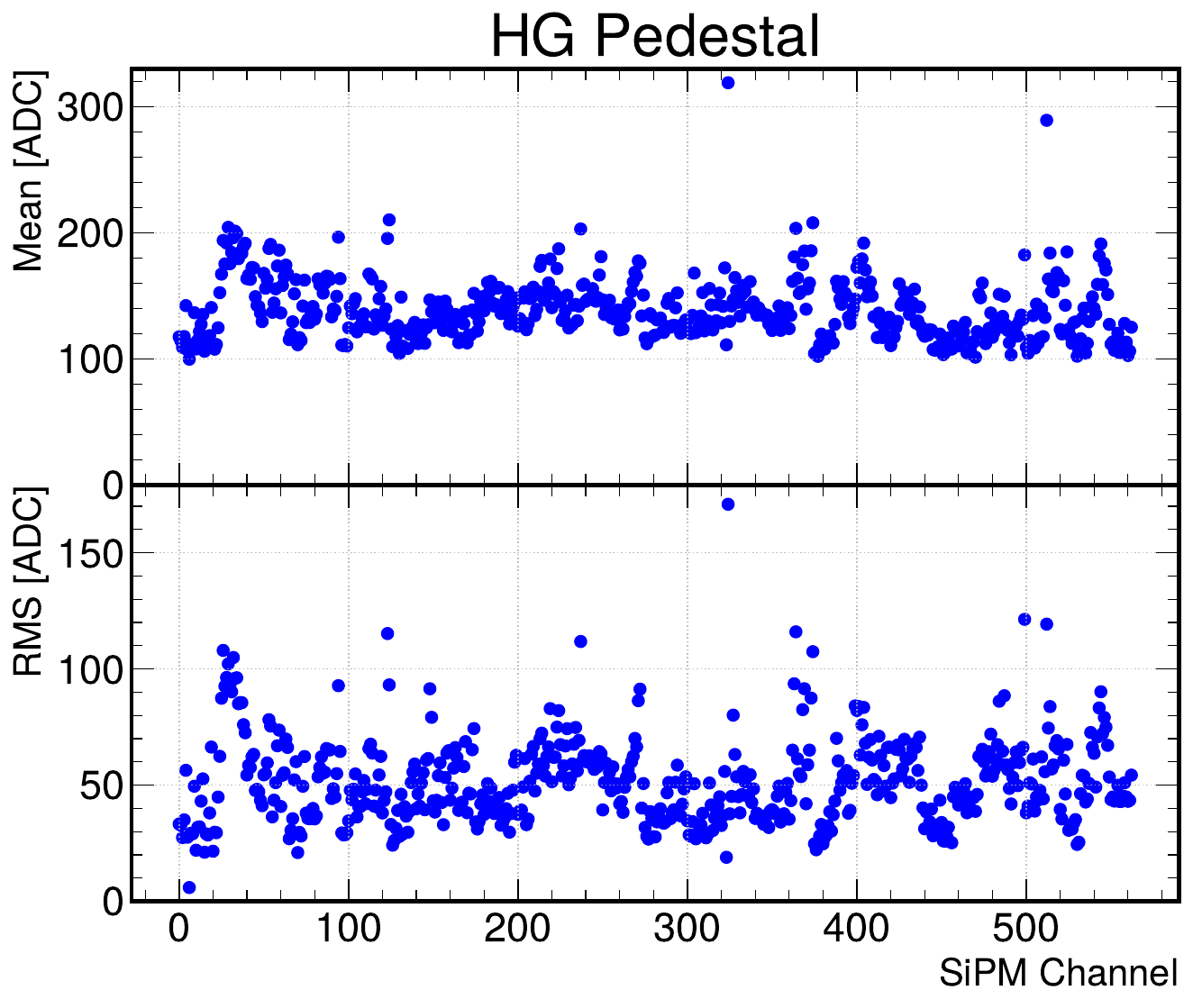}
    \includegraphics[width=0.45\linewidth]{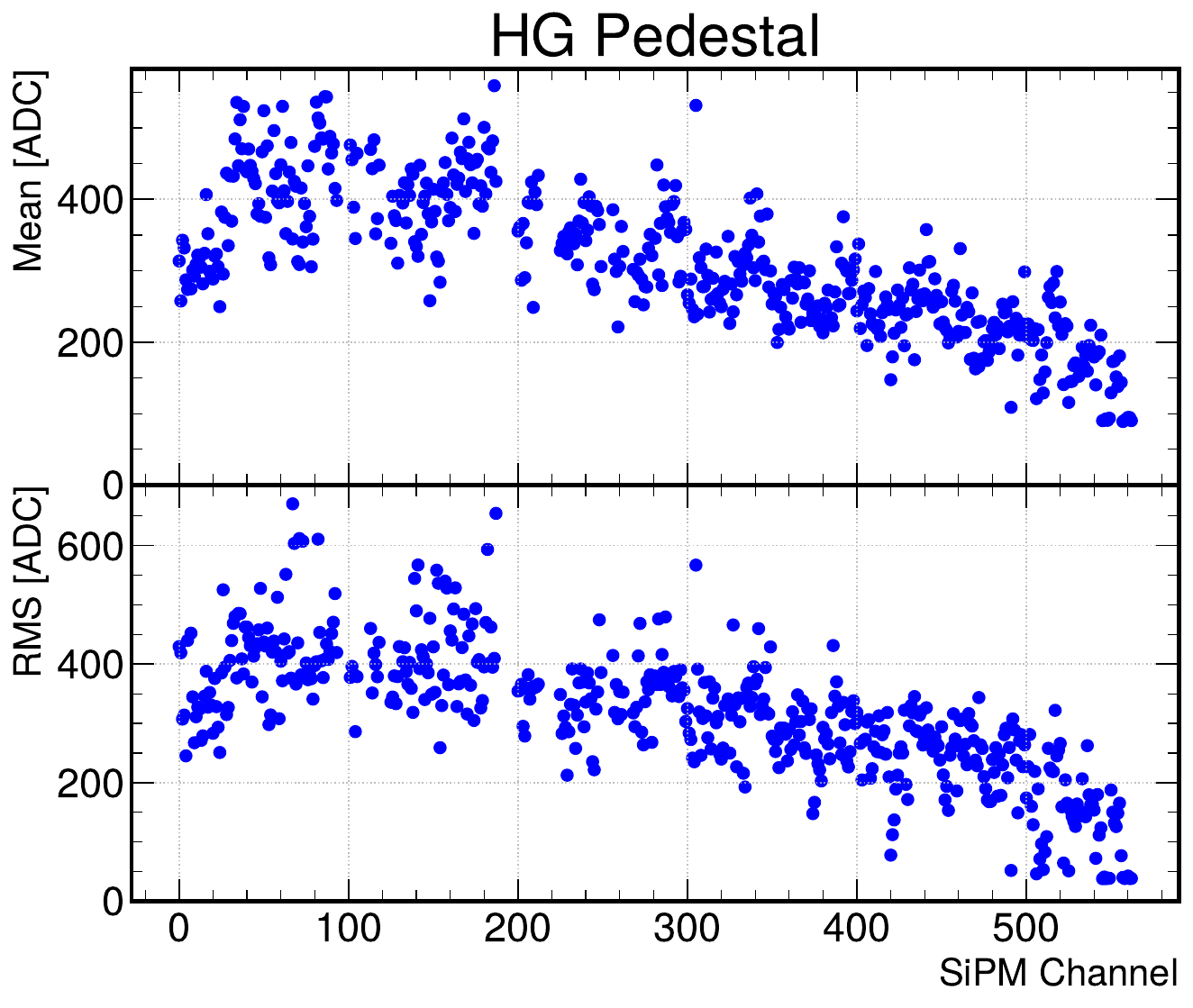}
    \caption{Representative pedestal means and widths across channels before (left) and after (right) irradiation.}
    \label{fig:pedestal}
\end{figure}

Before irradiation, the pedestal distributions were uniform across all channels, as illustrated in the left panel of figure~\ref{fig:pedestal}. The HG pedestal mean and width averaged around 150~ADCs and 50~ADCs, respectively, confirming stable baseline performance across the detector prior to irradiation.


After irradiation, several channels exhibited abnormal behavior: 32 became inactive, 4 showed unusually low pedestals, and 7 unusually high. The left panel of figure~\ref{fig:example_spectra_after_irradiation} shows a representative pedestal spectrum, featuring a markedly broader and shorter Gaussian distribution compared to pre-irradiation.


Excluding these problematic channels, the remaining channels revealed a clear layer-dependent trend, as seen in the right panel of figure~\ref{fig:pedestal}. Because the channel numbering increases with layer depth, the observed linearly decreasing pedestal mean and width with channel number reflects the longitudinal radiation gradient of the calorimeter. In the front layers, the pedestal mean rises significantly, approaching 500~ADCs with widths of a similar magnitude, while in the rear layers both the mean and width remain stable and consistent with pre-irradiation values.

This gradient along the calorimeter depth directly correlates with the non-uniform radiation profile observed in Sec.~\ref{subsec:results_radiation}. While pedestal shifts and broadening are significant in the irradiated front layers, the preserved stability in the rear layers ensures that channel-by-channel calibration remains viable.

\subsection{MIP Calibration}
\label{subsec:results_mip}

To evaluate the detector response after irradiation, the MIP calibration was repeated and compared to the pre-irradiation results. Only the high-gain (HG) mode is shown, since the low-gain spectra do not provide sufficient separation between the pedestal and the MIP peak.  

The right panel of figure~\ref{fig:example_spectra_after_irradiation} shows the MIP spectrum of a representative channel after irradiation. For reference, the right panel of figure~\ref{fig:example_spectra_before_irradiation} (pre-irradiation) exhibited a well-defined MIP peak, clearly separated from the pedestal. After irradiation, this separation is less pronounced: the pedestal distribution broadens significantly, resulting in a smeared and less distinct MIP peak.  

\begin{figure}[h!]
    \centering
    \includegraphics[width=0.45\linewidth]{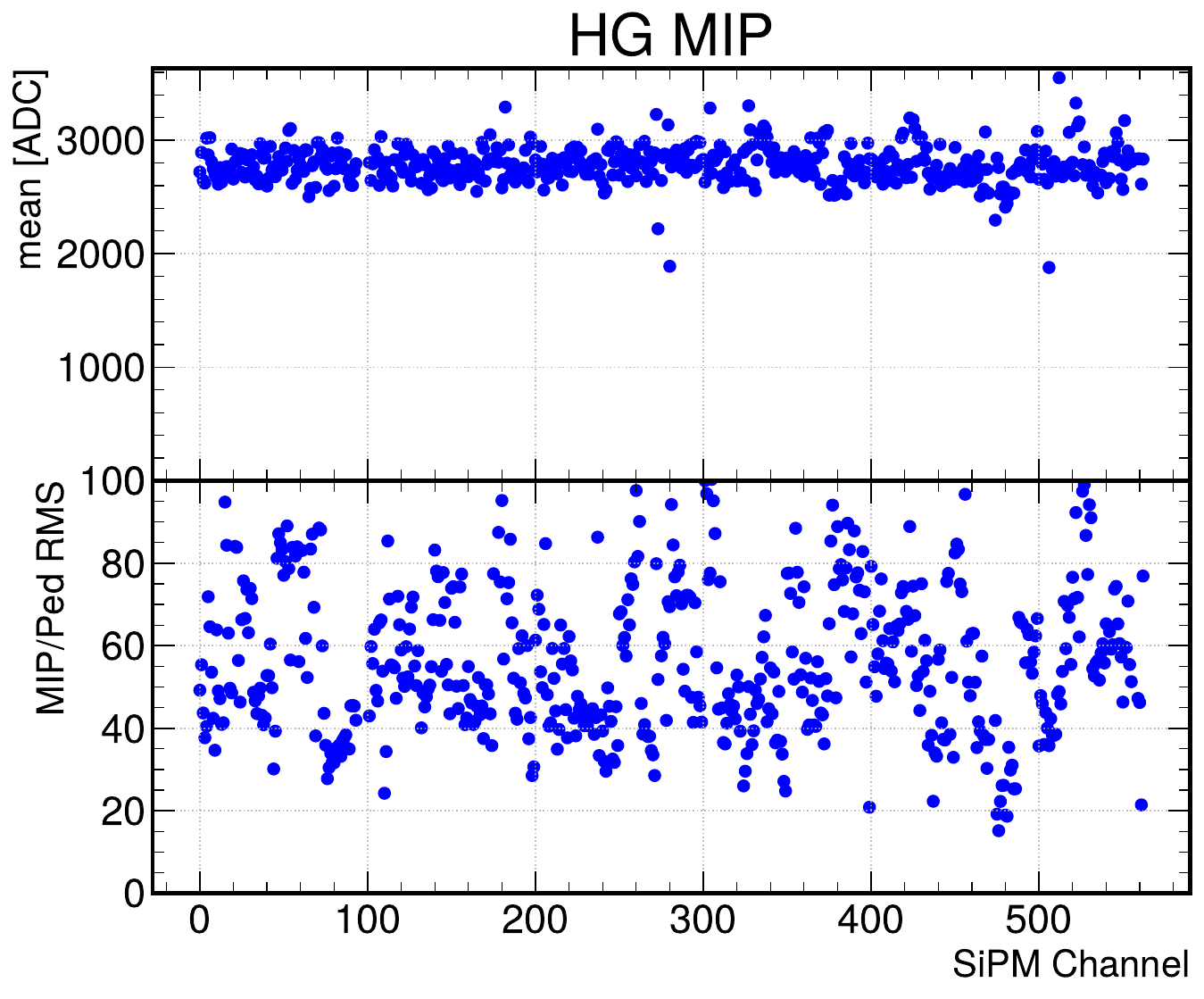}
    \includegraphics[width=0.45\linewidth]{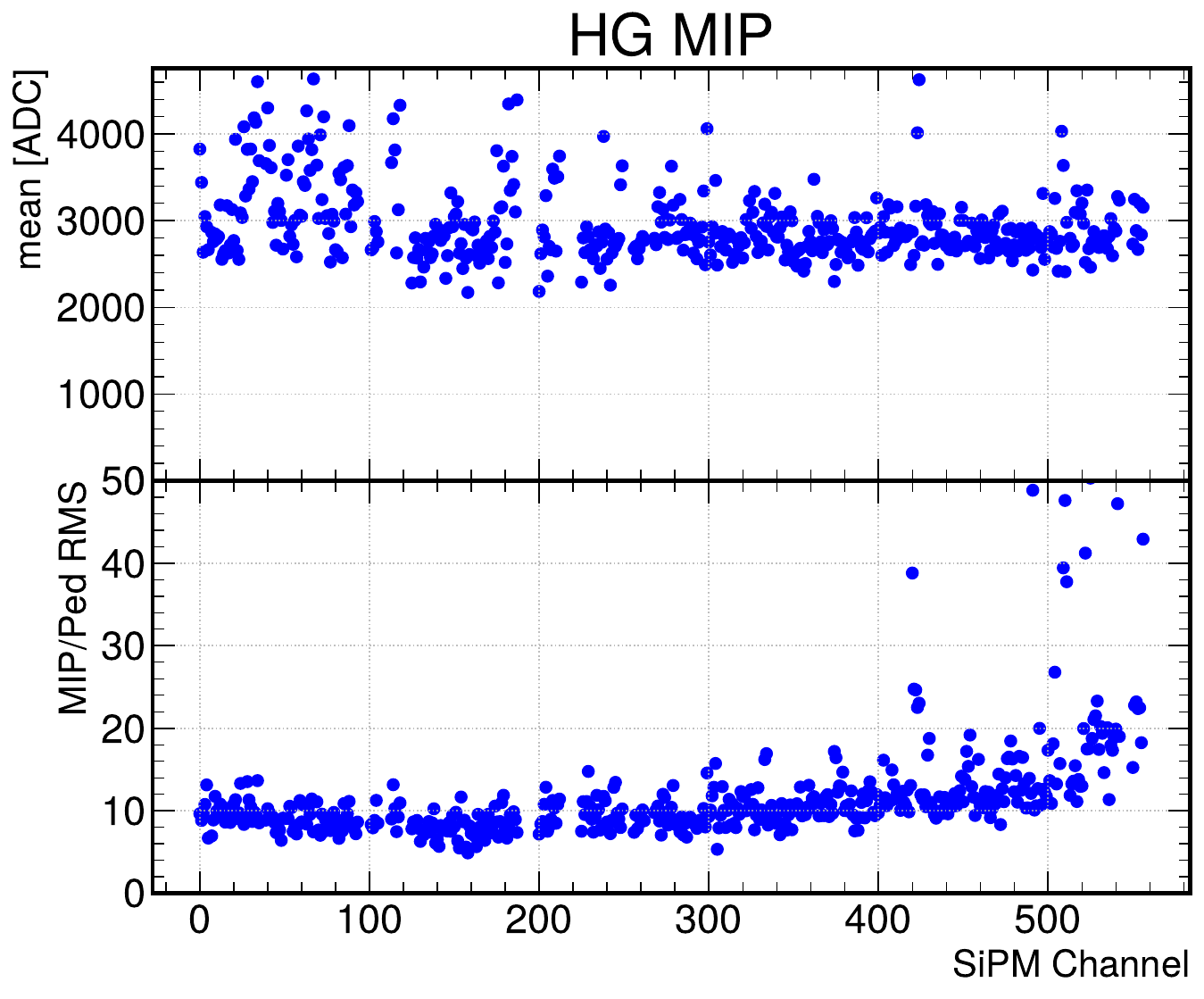}
    \caption{MIP calibration results for all channels before (left) and after (right) irradiation.}
    \label{fig:mip_all_channels}
\end{figure}

The channel-by-channel MIP calibration results are summarized in figure~\ref{fig:mip_all_channels}. Before irradiation, the MIP calibration values were uniform across channels, lying in the range of 2500--3000~ADCs. The ratio of MIP calibration to pedestal width exceeded 20 for all channels, with an average around 60, indicating excellent signal-to-noise performance.  

After irradiation, the calibration values increase to the range of 2500--4000~ADCs. In the front layers, where radiation damage is largest, the pedestal broadening substantially reduces the effective signal-to-noise ratio: the MIP-to-pedestal width ratio drops to about 5. In contrast, the rear layers remain close to their pre-irradiation performance.  

In summary, irradiation leads to higher calibration values and broader pedestal distributions, with the strongest impact observed in the front layers. Nonetheless, the post-irradiation signal-to-noise ratio remains above 5 across all layers, ensuring that the detector can still be effectively calibrated and reliably operated.


\section{Discussion and Conclusions}
\label{sec:conclusion}

We have presented a systematic study of radiation effects on SiPM-based calorimeter prototypes intended for the ZDC of the ePIC detector. The prototype was exposed to controlled proton irradiation at NSRL, with radiation fluence levels chosen to emulate the annual dose expected in ePIC under nominal luminosity running. By combining post-irradiation dark current measurements with benchmark curves from previous studies, we established that the prototype received fluences up to $10^{10.98}$~1-MeV~$n_\text{eq}$/cm$^2$ in the most exposed layers, closely matching the design expectations for the ZDC front section after one year of EIC operation.  

Pedestal studies revealed that, while a few channels became inactive or exhibited abnormal response after irradiation, the majority of channels remained operational. Excluding the problematic channels, the pedestal mean and width showed a clear layer dependence, reflecting the longitudinal radiation profile of the calorimeter. In the most irradiated front layers, pedestal broadening was pronounced, with widths increasing by a factor of three in LG and nearly an order of magnitude in HG. The rear layers, however, retained pre-irradiation performance, demonstrating that radiation damage is localized in accordance with dose gradients.  

MIP calibration results showed that the detector still preserves a clear minimum-ionizing particle response after irradiation. Although pedestal broadening reduced the effective signal-to-noise ratio in the front layers from $\sim$60 to $\sim$5, this remains above the threshold required for reliable calibration and operation. Importantly, the rear layers preserved both calibration values and high signal-to-noise ratios, ensuring uniform performance throughout most of the detector depth.  

Overall, these results demonstrate that the SiPM-on-tile calorimeter technology can withstand radiation levels equivalent to one year of nominal ePIC ZDC operation, while still providing sufficient pedestal stability and MIP calibration fidelity for physics measurements. The identification of non-functioning and anomalous channels highlights potential failure modes that merit further investigation. Future studies will include extended irradiation campaigns to probe multi-year exposures, as well as complementary annealing and recovery tests to better understand long-term operational stability.  

It should be noted that the current tests were conducted using the CITIROC-1A ASIC for readout. Once the EIC-custom CALOROC ASIC becomes available, the readout system will be updated accordingly, while similar performance is expected and the current results provide a first step toward validating the full design.

In conclusion, the prototype tests provide strong evidence that the chosen technology is viable for the ZDC at ePIC, with radiation-induced performance degradation remaining manageable within the expected operational conditions.

\acknowledgments

We would like to sincerely thank Michael Sivertz and Xiaodong Jiang, as well as all the staff at the NASA Space Radiation Laboratory (NSRL), for their invaluable support; without their assistance, the test beam would not have been possible. We also thank Sebouh J. Paul for creating the beautiful schematic plot included in this paper. Sean Preins was supported by a HEPCAT fellowship from DOE award DE-SC0022313.


\bibliographystyle{JHEP}
\bibliography{biblio.bib}


%
%
%
%
\end{document}